\documentclass[seceq]{ptptex}
\usepackage{graphicx}
\usepackage{subfigure}

\begin{document}

\arraycolsep1.5pt

\newcommand{\Ima}{\textrm{Im}}
\newcommand{\Rea}{\textrm{Re}}
\newcommand{\mev}{\textrm{ MeV}}
\newcommand{\be}{\begin{equation}}
\newcommand{\ee}{\end{equation}}
\newcommand{\ba}{\begin{eqnarray}}
\newcommand{\ea}{\end{eqnarray}}
\newcommand{\gev}{\textrm{ GeV}}
\newcommand{\nn}{{\nonumber}}

\markboth{
D.~Cabrera, D.~Jido, R.~Rapp and L.~Roca%
}{
The $a_1(1260)$ as a $\rho\pi$ resonance in nuclear matter%
}

\title{The $a_1(1260)$ as a $\rho\pi$ resonance in nuclear matter}

\author{D.~\textsc{Cabrera}$^{1}$, D.~\textsc{Jido}$^{2}$, 
R.~\textsc{Rapp}$^{3}$ and L.~\textsc{Roca}$^{4,2}$}
\inst{
$^{1}$Departamento de F\'{\i}sica Te\'orica II, Universidad Complutense,
28040 Madrid, Spain \\
$^{2}$Yukawa Institute for Theoretical Physics, Kyoto
University, Kyoto 606--8502, Japan \\
$^{3}$Cyclotron Institute and Physics Department, Texas A\&M
University, College Station, Texas 77843-3366, U.S.A. \\
$^{4}$Departamento de F\'{\i}sica, Universidad de Murcia, E-30071 Murcia, Spain}


\date{\today}

\abst{
We present a theoretical study of the properties of the  $a_1(1260)$
axial-vector resonance in a cold nuclear medium. In the vacuum, the 
$a_1(1260)$ resonance is generated dynamically from the interactions 
of a pseudoscalar and vector meson ($\rho\pi$ and $K\bar{K}^*$) 
in a coupled channel chiral unitary approach. Medium effects are
implemented  through the modification of the $\rho$ and $\pi$
propagators at finite nuclear density from well established microscopic
many-body calculations. The in-medium pion spectral function accounts
for the coupling to $N$-hole and $\Delta$-hole excitations including
short range correlations, whereas the in-medium $\rho$ incorporates
modifications of its virtual pion cloud as well as direct
resonance-hole excitations. The resulting in-medium $a_1(1260)$ 
exhibits significant broadening with increasing density as reflected 
in the $\rho \pi$ scattering amplitude. The possible
relation of our results with partial restoration of chiral symmetry
in nuclear matter is discussed in terms of in-medium Weinberg sum rules.
}

\maketitle

\section{Introduction}
\label{Intro}

The investigation of modifications of  hadron properties in nuclear matter
has been a vigorous activity in nuclear physics over the past $\sim$30 
years. The origin of the interest is at least twofold. On the one hand, 
the modification of the in-medium properties of hadrons has direct impact
on the nuclear equation of state, as relevant, for instance, 
in the description of nuclear saturation~\cite{Rapp:1997ii}, 
neutron stars~\cite{Weber:2007ch} and low-energy heavy-ion 
collisions~\cite{Li:2008gp}.
On the other hand, it is expected that in-medium hadronic spectral 
functions encode precursor effects of transitions to 
new phases of nuclear matter with modified symmetry properties, such 
as the partial restoration of the chiral symmetry of QCD in nuclear 
matter~\cite{Hayano:2008vn,Rapp:2009yu}.
In either case, quantitative evaluations of hadronic spectral
properties in a consistent framework are mandatory to realize 
the pertinent objectives. Large-scale experimental efforts are devoted 
to search for in-medium modifications especially of mesonic excitations, 
including pion and dilepton production experiments off nuclear 
targets~\cite{Starostin:2000cb,Grion:2005hu,Metag:2007zz,Djalali:2007zz},
and dilepton production in heavy-ion 
collisions~\cite{Adamova:2006nu,Pachmayer:2008yn,Arnaldi:2008fw}.
In particular in the latter, which are mostly sensitive to the in-medium
$\rho$ spectral function, the produced hot and dense QCD medium is 
expected to evolve to a significant extent in the vicinity of the chiral 
restoration transition. Thus, to properly interpret the $\rho$ signal,
it is essential to establish its theoretical connection to chiral symmetry.  

In the standard interpretation of chiral symmetry, the chiral partner
of the vector-isovector current (dominated by the $\rho$ meson) is 
identified with the axialvector-isovector one (dominated by the $a_1$,
plus a derivative of the pion current)\footnote{See, however, 
Ref.~\citen{Harada:2003jx} for an alternative interpretation in terms
of the so-called vector manifestation.}.
Therefore, if chiral symmetry is restored in the nuclear medium, there
should be no difference between left-handed and right-handed currents
or, in other words, between vector and axialvector current-current 
correlators (and spectral functions).
This elevates the study of the modification of the axialvector
channel to an important theoretical issue, even if its experimental
measurement will not turn out to be feasible. 
Such studies will be helpful in discriminating scenarios of chiral 
symmetry restoration, which, in principle, can be rather different. 
E.g., two extreme scenarios
for the degeneration of vector and axialvector spectral functions are 
(i) the vector and axialvector mesons survive as well defined states
with a common mass;
(ii) both vector and axialvector spectral functions are strongly
broadened dissolving any quasiparticle poles (``melting resonance'' 
scenario~\cite{Rapp:1999us}).

Thus far relatively few calculations to evaluate the properties of 
the axialvector resonances in the medium have been performed. 
In Ref.~\citen{Kim:1999pb} a schematic ``$a_1$-sobar"
model at finite density has been constructed where an important role is 
played by the putative chiral partners of the $\rho$-sobar 
resonance-hole excitations (e.g., $N(1520)N^{-1}$, etc.); see also
Ref.~\citen{Rapp:2002pn}.
In Ref.~\citen{Urban:2001ru}, a linear $\sigma$ model with a global 
implementation of vector mesons into the chiral pion Lagrangian (as 
opposed to local ones in the Hidden-Local-Symmetry (HLS) and 
Massive-Yang-Mills (MYM) approaches) has been constructed and applied 
at finite temperature.  In Ref.~\citen{Harada:2008hj} finite-temperature 
studies have been carried out based on different local realizations of 
chiral symmetry within the HLS scheme. 
A gauged linear $\sigma$ model was studied in a selfconsistent mean-field
scheme at finite temperature in Ref.~\citen{Struber:2007bm}; see also
Ref.~\citen{Pisarski:1995xu}. 
In all of these studies the $a_1$ meson was introduced
as a genuine degree of freedom. 

In the last few years several works~\cite{Lutz:2003fm,Roca:2005nm,Geng:2006yb}  
have reported arguments and evidence for a dynamical nature of low-lying 
axialvector resonances in vacuum, utilizing variants of the chiral unitary
approach~\cite{Dobado:1996ps,Oller:1997ti,Oller:1998hw,Oller:1999zr,Kaiser:1995eg,Oller:2000fj,Lutz:2001yb,Geng:2008ag}.
The latter has been widely and successfully applied to various  
meson-meson~\cite{Oller:1997ti,Kaiser:1998fi,Oller:1998hw,Oller:1999zr,Nieves:2000bx,Gamermann:2006nm}
and
meson-baryon~\cite{Kaiser:1995eg,Oset:1997it,Oller:2000fj,GarciaRecio:2003ks,Hyodo:2002pk,Jido:2003cb,Oller:2006jw,Sarkar:2004jh,Roca:2008kr} 
systems, both in vacuum and in cold nuclear matter. 
One of the main results is that many meson and baryon resonances can
be described by meson-meson or meson-baryon dynamics without introducing
a genuine field into the hadronic Lagrangian\footnote{Recently 
a method to test whether dynamically described states can be interpreted as
quasibound states of hadrons has been proposed in Ref.~\citen{Hyodo:2008xr}.}. 
Also the 
axialvector resonances naturally appear \cite{Lutz:2003fm,Roca:2005nm} as
poles in the scattering matrix of the interaction of pseudoscalar mesons with
vector mesons, once coupled channels with the only input of the lowest order 
chiral Lagrangian are accounted for (including the lowest-lying vector
mesons, but without explicit axialvector fields). The low-lying axialvector 
can be described by dynamics of a pseudoscalar and a vector meson. 
In particular, the $a_1(1260)$ can be interpreted as a strongly correlated
$\rho\pi$ system.  

In the present work, we combine the picture of a dynamically generated 
axialvector resonance~\cite{Roca:2005nm} with hadronic many-body 
techniques, to obtain an estimate of the in-medium axialvector spectral 
function. An advantage of this approach is that one does not
have to specify scarcely known couplings of the $a_1$ to higher
resonances (both in scattering off pions at finite temperature and off 
nucleons at finite density).  
The medium modifications are implemented in the intermediate unitary
loops, specifically via in-medium $\pi$ and $\rho$ spectral functions
which are rather well known from detailed many-body calculations 
in connection with comparisons to a wide range of data (both in nuclei
and in heavy-ion reactions). 
Since our model for the (generated) $a_1(1260)$ is still beset
with significant uncertainties even in the vacuum (mainly due to the
large $a_1$ width)~\cite{Roca:2005nm}, it is not yet warranted to
aim at a complete assessment of all possible in-medium effects. 
Based on previous experience in other meson-meson and meson-baryon systems, 
we will therefore restrict ourselves to the in-medium renormalization of 
intermediate pseudoscalar and vector-meson states, which should provide
an estimate of the most prominent medium modifications.

Our article is organized as follows: in Sec.~\ref{sec:modelvacuum} we 
summarize the formalism of pseudoscalar-vector meson interactions in 
vacuum, which is the starting point for the present calculation, and 
emphasize its main virtues and uncertainties.
The inputs for the pion and $\rho$-meson self-energies in cold nuclear 
matter are discussed in Sec.~\ref{sec:nucmed}, which we use to describe 
in some detail how the $\rho\pi$ 2-particle propagator is modified at 
finite density.
Sec.~\ref{sec:results} contains our results for the in-medium $\rho \pi$
scattering amplitude (specifically around the $a_1(1260)$ pole), as well 
as an estimate
for its impact on partial chiral symmetry restoration by evaluating
in-medium Weinberg sum rules. In Sec.~\ref{sec:concl} we draw conclusions 
and give on outlook to future lines of investigation.

\section{The pseudoscalar-vector meson interaction in vacuum}
\label{sec:modelvacuum}
We start by introducing our model for the $a_1(1260)$ resonance 
in free space. In Ref.~\citen{Roca:2005nm} several axialvector 
resonances are dynamically generated with the only input of the 
lowest-order chiral Lagrangian and unitarity in coupled channels. 
Most of the experimentally known low-lying axialvector resonances can 
be accounted for as emerging from the interaction of a vector ($V$) and 
a pseudoscalar ($P$) meson, manifesting themselves as
poles in unphysical Riemann sheets of the $VP$ scattering amplitudes.

Considering the vector mesons as
fields transforming homogeneously under the nonlinear
realization of chiral symmetry \cite{WCCWZ}, 
the interaction of two vector
and two pseudoscalar mesons at lowest order in the pseudoscalar
fields can be obtained from the following interaction Lagrangian 
\cite{Birse:1996hd}: 
\be
\label{eq:lag} {\cal
L}=-\frac{1}{4}\{(\nabla_\mu V_\nu-\nabla_\nu V_\mu)(\nabla^\mu
V^\nu-\nabla_\mu V_\nu) \} \ , \quad 
\ee 
where $\nabla_\mu V_\nu=\partial_\mu V_\nu+[\Gamma_\mu,V_\nu]$
is the $SU$(3)-matrix valued covariant derivative, 
with the $SU$(3) connection defined as
$\Gamma_\mu=(u^\dagger \partial_\mu u+u\partial_\mu u^\dagger)/2$,
$u=\exp(P/\sqrt{2} f)$ and
\begin{eqnarray}
&&P\equiv\left(
\begin{array}{ccc}
\frac{\pi^0}{\sqrt{2}}+\frac{\eta_8}{\sqrt{6}}& \pi^+& K^+ \\
\pi^-&-\frac{\pi^0}{\sqrt{2}}+\frac{\eta_8}{\sqrt{6}}& K^0\\
K^-& \bar{K}^0&-\frac{2\eta_8}{\sqrt{6}}
\end{array}
\right) \ ,
\;\\ \nonumber
&&V_\mu\equiv\left(
\begin{array}{ccc}
\frac{\rho^0}{\sqrt{2}}+\frac{\omega}{\sqrt{2}}& \rho^+& K^{*+} \\
\rho^-&-\frac{\rho^0}{\sqrt{2}}+\frac{\omega}{\sqrt{2}}& K^{*0}\\
K^{*-}& \bar{K}^0&\phi
\end{array}
\right)_\mu \ .
\end{eqnarray}
The Lagrangian of Eq.~(\ref{eq:lag}) is invariant under 
chiral transformations $SU(3)_L\otimes SU(3)_R$.

For the present work we only need the interaction terms containing 
two vector and two pseudoscalar mesons. Thus,
expanding the Lagrangian of Eq.~(\ref{eq:lag}) up to two vector and
two pseudoscalar fields one obtains \cite{Lutz:2003fm,Roca:2005nm}
\be\label{eq:lag2} 
{\cal L}_{VP}=-\frac{1}{4f^2}\langle[V^\mu,\partial^\nu V_\mu]
[P,\partial_\mu P]\rangle \ , \quad
\ee
which allows to evaluate the $VP\to VP$ tree level amplitudes. Their 
explicit expression, after projecting onto $S$-waves, reads
\begin{equation}
\epsilon\cdot\epsilon' \
V_{ij}(s)=-\frac{\epsilon\cdot\epsilon'}{8f^2} C_{ij}
\left[3s-(M^2+m^2+M'\,^2+m'\,^2)
-\frac{1}{s}(M^2-m^2)(M'\,^2-m'\,^2)\right] \ ,
\label{eq:Vtree}
\end{equation}
where $\epsilon$ ($\epsilon'$) denotes the polarization four-vector of the
incoming (outgoing) vector meson. The masses $M (M')$ and $m (m')$ correspond 
to the initial (final) vector mesons and initial (final) pseudoscalar mesons,
respectively, and we use an average value for each isospin multiplet. The
indices $i$ and $j$ represent the initial and final $VP$ states, respectively,
in the isospin basis. The numerical coefficients, $C_{ij}$, can be 
fixed by the $SU(3)$ flavor symmetry and the values can be
found in Ref.~\citen{Roca:2005nm}. For the present study of the $a_1(1260)$,  
we have only two channels; the
$\rho\pi$ channel and the negative $G$-parity combination $1/\sqrt{2}(|\bar
K^* K>-|K^*\bar K>)$. 
Note that the tree level amplitude in
Eq.~(\ref{eq:Vtree}) only depends on the pion decay constant, $f=93\mev$.

\begin{figure}[!t]
\begin{center}
\includegraphics[width=0.8\textwidth]{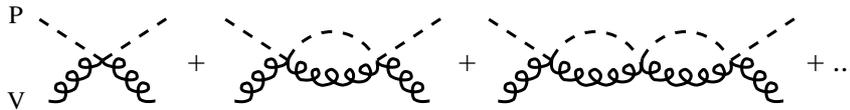}
\caption{Diagrammatic interpretation of
 the unitarization of the $VP\to VP$ amplitude.}
\label{fig:bethe}
\end{center}
\end{figure}

The full $T$-matrix can now be obtained by unitarizing the tree level 
amplitudes from the chiral Lagrangian, cf.~Eq.~(\ref{eq:Vtree}). 
In Ref.~\citen{Roca:2005nm} this was carried out within a coupled channel 
Bethe-Salpeter formalism (equivalent to the Inverse Amplitude 
Method~\cite{Dobado:1996ps,Oller:1998hw} or the $N/D$ 
method~\cite{Oller:1999zr}) for the two-meson interaction. 
The transverse part of the $VP\to VP$ unitarized scattering amplitude, 
which encodes possibly dynamically generated resonance poles (see 
Ref.~\citen{Roca:2005nm} for details), takes the form 
\be
\label{eq:bethe} 
T=-[1+V G]^{-1}V\,\vec \epsilon \cdot \vec \epsilon' \ , 
\ee
where the $V$-matrix
elements are those of Eq.~(\ref{eq:Vtree}).
The matrix $G$ 
is diagonal with the $l^{\rm th}$
element, $G_l$, 
given by the two-meson loop function,
\be
\label{eq:G}
G_l(P)=i\,\int\frac{d^4 q}{(2\pi)^4} \,
\frac{1}{(P-q)^2-M_l^2+i\epsilon}\,\frac{1}{q^2-m_l^2+i\epsilon}
\ ,
\ee
where $P$ is the total four-momentum, $P^2=s$. 
In the $\rho\pi$ channel, which is the main
channel in building up the $a_1(1260)$, the width of the
$\rho$ meson is significant ($\simeq 150\,\mev$).  This has 
been considered in the evaluation of the $G_{\rho\pi}$ function by
replacing the stable-particle vector meson propagator in
Eq.~(\ref{eq:G}) by the appropriate version accounting for the 
two-pion decay width (or
self-energy). This will be discussed in more detail in the next section.
Eq.~(\ref{eq:bethe}) represents the resummation
of the diagrammatic series shown in Fig.~\ref{fig:bethe}.

The energy-momentum integration in Eq.~(\ref{eq:G}) for $G_l$ is 
divergent and hence has to be regularized. This was done in 
Ref.~\citen{Roca:2005nm} in either a cut-off or dimensional 
regularization scheme. Both methods lead to qualitatively similar 
results.  In the present study we will use a cut-off in the 
three-momentum of the intermediate particles, which provides a 
transparent regularization method in a nuclear medium.
The regularization procedure introduces the only free parameter of
the model, $q_{\rm{max}}$, which we
take of a ``natural" size, $q_{\rm max}\sim 1\gev$.

The $T$-matrix obtained from Eq.~(\ref{eq:bethe}) contains the information 
on possible resonances dynamically generated from the $VP$ interaction. 
The identification of the resonances and 
the assignment of their main properties, i.e., mass and width, is 
not unique. In scattering theory it is common practice to identify 
resonances with poles on unphysical Riemann sheets of a certain 
partial-wave amplitude. In Ref.~\citen{Roca:2005nm}, several poles 
of the $T$-matrix in the second Riemann sheet were found which have 
been associated with the $a_1(1260)$, $b_1(1235)$, $h_1(1170)$, 
$h_1(1380)$, $f_1(1285)$ and the two $K_1(1270)$ resonances. If
the pole position is not very far from the real axis and there are no
thresholds close by, the real and imaginary parts of the pole position 
approximately define the mass and half-width of the resonance. For the
quantum numbers of the $a_1(1260)$ channel we find a pole at 
$\sqrt{s_{\rm pole}}=(1117-{\rm i}\,177)$~MeV. 
This would imply a width of the order of
$350\mev$. Note that in the PDG the mass is given as 
$M_{a_1}=(1230\pm40)\mev$ and the value for the width is quoted with
large uncertainty,
$\Gamma_{a_1}\sim$\,250-600~MeV.  The unitarized $VP$ scattering
amplitude discussed here has been recently used in a detailed  analysis of 
the $\tau \to \pi\pi\pi \nu$ decay 
spectrum~\cite{Wagner:2007wy,Wagner:2008gz} and compared to experimental 
data from the ALEPH~\cite{Schael:2005am} and CLEO~\cite{Asner:1999kj} 
collaborations. The model reproduces the data well with a
single regularization parameter, a momentum cut-off of the same size as used
in Ref.~\citen{Roca:2005nm} and in the present work, and without the need of
introducing an explicit $a_1$ field in the theory. In fact, introducing
an explicit $a_1$ was found to worsen the agreement with experiment unless 
an ``unnatural" choice for the cut-off value is made. The results from
Refs.~\citen{Wagner:2007wy,Wagner:2008gz} provide a phenomenological test of
our dynamically generated $a_1(1260)$ and increase confidence in the model
as a starting point for our calculation in nuclear matter.

As discussed in Ref.~\citen{Roca:2005nm}, we assign to our calculated $a_1$
pole position an uncertainty of about $100\mev$ for the real part and of
about $50$\% for the imaginary part, based on different
regularization schemes and the consideration of a finite decay width for the
intermediate vector meson. This may not be of too great a concern given 
that the resonance is far from the real axis and that 
the experimental width has a large uncertainty. 
The experimentally accessible information, which is also well defined 
from a theoretical point of view, is the scattering amplitude on the real 
energy axis. Hence, we will base the discussion of our results largely 
on the $\rho\pi$ amplitude and its modification due to a finite nuclear 
density. The estimated values of the mass and decay width of the
resonance (extracted from the pole position) are not of much relevance in 
the present work since we are  interested in the relative change in the
nuclear medium with respect to the free case. Our aim is to provide an
estimation of the modifications that the $a_1(1260)$ experiences in the
nuclear medium as a consequence of many-body dynamics in the intermediate 
$\rho\pi$ system.

\section{Nuclear Medium Modifications}
\label{sec:nucmed}
As discussed above, the nuclear medium effects on the dynamically generated
$a_1$ will be restricted to many-body calculations of the $\rho$ 
and $\pi$ self-energies. We neglect medium effects on the hidden strangeness 
channel as it lies higher in energy and has a weaker coupling to the 
$a_1$ spectrum, as we shall see in the next section.

In addition to the modified meson propagators, in-medium vertex corrections 
to the vacuum $VP$ interaction, in principle, occur. Vertex corrections 
play a crucial role, for instance, in restoring transversality of the 
$\rho$-self-energy tensor in nuclear 
matter~\cite{Chanfray:1993ue,Herrmann:1993za,Urban:1998eg}. Another 
example are scalar-isoscalar $\pi\pi$ correlations in nuclear matter, 
where vertex corrections associated with the $\pi N$ interaction are 
required by chiral symmetry~\cite{Chiang:1997di}. In the present model, 
we expect vertex corrections to be necessary in connection with the partial 
conservation of the axial current (PCAC), which should be maintained
in the presence of a nuclear medium (especially in scenarios
where the system approaches chiral symmetry restoration). The manifestation
of PCAC at the level of the $\rho \pi$ amplitude in our model is not 
obvious, since correlated $VP$ exchange only accounts for the 
(low-energy) resonant part of the axialvector correlator (additional
non-resonant (or background) terms, such as uncorrelated $3\pi$ exchange, 
have to be included at the same level~\cite{Wagner:2008gz}). Including 
vertex corrections in the present model would thus require a
systematic analysis of the Lagrangian terms involving pseudoscalar, 
vector meson and baryon interactions which is beyond the scope of this
paper. It shall be addressed in future work within an evaluation of the 
full axialvector spectral function in nuclear matter. Generally, we
expect that in-medium vertex correction enhance the medium
effects on the $\pi\rho$ amplitude and axialvector correlator.

\begin{figure}[!t]
\begin{center}
\includegraphics[width=0.5\textwidth]{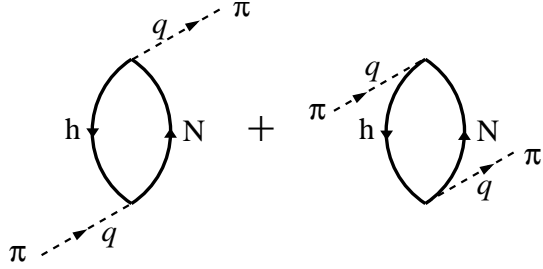}
\caption{Particle-hole ($Nh$) excitation diagrams contributing to the $P$-wave
pion self-energy.}
\label{fig:ph-diagram}
\end{center}
\end{figure}
Let us turn to the in-medium pion propagator.
In cold nuclear matter, the pion spectrum
exhibits a mixture of the pion quasi-particle mode and predominantly 
$P$-wave nucleon-hole ($Nh$) and Delta-hole ($\Delta h$) 
excitations~\cite{Oset:1981ih,Migdal:1990vm}, as depicted diagrammatically 
in Fig.~\ref{fig:ph-diagram} for the $s$- and $u$-channel $Nh$ excitation.
The lowest-order irreducible $P$-wave pion self-energy reads
\begin{eqnarray}
\label{eq:piself-total}
\Pi^p_{\pi} (q^0,\vec{q};\varrho) =
\frac{\left( \frac{f_N}{m_{\pi}} \right) ^2
F_{\pi}(\vec{q}\,^2) \, \vec{q}\,^2 \,
\left[ U_{NN^{-1}} (q^0,\vec{q};\varrho) + U_{\Delta N^{-1}} 
(q^0,\vec{q};\varrho) \right]}
{1 - \left( \frac{f_N}{m_{\pi}} \right) ^2 \, g' \,
\left[ U_{NN^{-1}} (q^0,\vec{q};\varrho) + 
U_{\Delta N^{-1}}(q^0,\vec{q};\varrho)
 \right]}
\,\,\, ,
\end{eqnarray}
where $U$ denotes the Lindhard function at a nuclear density 
$\varrho$~\cite{Oset:1989ey}, and $f_N$ ($f_{\Delta}$) is the $\pi NN$ ($\pi
N \Delta$) coupling constant determined from analyses of pion-nucleon and 
pion-nucleus reactions, 
$f_N \simeq 1$ and $f_{\Delta}/f_N \simeq 2.13$ (a factor of 
$f_{\Delta}/f_N$ is absorbed in the definition of the $\Delta h$
Lindhard function). The strength of the collective modes of the pion is
modified by repulsive, spin-isospin $NN$ and $N\Delta$ short-range 
correlations~\cite{Oset:1981ih}. 
In Eq.~(\ref{eq:piself-total}) the latter are accounted for 
in a phenomenological way
with a single Landau-Migdal interaction parameter, $g'=0.7$, and the
corresponding RPA series for the pion self-energy is resummed 
(somewhat different 
couplings for $NN$-$N\Delta $ and $N\Delta$-$N\Delta$ interactions 
have been considered, for instance, in Ref.~\citen{Urban:1998eg}). 
Eq.~(\ref{eq:piself-total}) includes finite-size effects on the $\pi NN$ 
and $\pi N \Delta$ vertices via hadronic monopole form-factors, 
\be
F_{\pi}(\vec{q}\,^2) = \Lambda_{\pi}^2 /
(\Lambda_{\pi}^2 + \vec{q}\,^2) \ ,
\label{eq:fformfact}
\ee
with $\Lambda_{\pi}\sim 1$~GeV.

A more general parameterization of the short-range correlation effects in the
pion self-energy is given in Ref.~\citen{Urban:1998eg}, in terms of three
different Migdal parameters in a coupled-channel set-up:
 $g'_{11}$ ($NNNN$ vertex),
$g'_{12}$ ($NNN\Delta$ vertex) and $g'_{22}$ ($NN\Delta\Delta$ vertex). One 
obtains for the pion self-energy
\ba
\label{eq:piself-total-rapp}
\lefteqn{\Pi^p_{\pi} (q^0,\vec{q};\varrho) =} && \nonumber \\ &&
\frac{\left( \frac{f_N}{m_{\pi}} \right) ^2
F_{\pi}(\vec{q}\,^2) \, \vec{q}\,^2 \,
\left[ U_{NN^{-1}}  + U_{\Delta N^{-1}}
-(g'_{11}-2g'_{12}+g'_{22})U_{NN^{-1}} U_{\Delta N^{-1}}
\right]}
{1 - \left( \frac{f_N}{m_{\pi}} \right) ^2 \, 
\left[ g'_{11}U_{NN^{-1}} 
+ g'_{22}U_{\Delta N^{-1}} 
-(g'_{11}g'_{22}-g_{12}'^2) U_{NN^{-1}} 
U_{\Delta N^{-1}}
 \right]}
\,\,\, .
\ea
We shall adopt the parameter values in \cite{Urban:1998eg}, namely,
$g'_{11}=0.6$, $g'_{12}=g'_{22}=0.2$ and
$\Lambda_{\pi}=0.3$~GeV, which were obtained from an analysis of 
pion-induced vector-meson production, $\pi N\to\rho
N$ \cite{Rapp:1999ej}.
In the following we will denote as $\pi$-self-energy B the parameterization 
in Eq.~(\ref{eq:piself-total-rapp}) with the parameter values quoted above, 
and refer to Eq.~(\ref{eq:piself-total}) as $\pi$-self-energy A with the 
parameter values quoted thereafter. 
Using both models A and B for the pion self-energy will give some
indication of the uncertainties in our final results for the properties 
of the $a_1$ resonance in a nuclear medium.

\begin{figure}[!t]
\begin{center}
\includegraphics[width=0.8\textwidth]{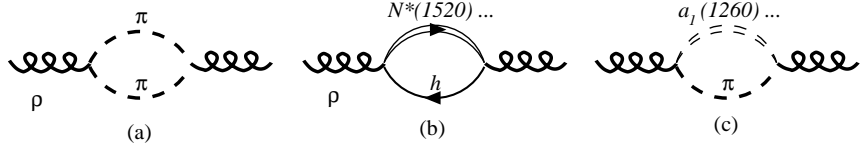}
\caption{Contributions to the $\rho$-meson self-energy in nuclear matter
from the coupling to (a) the two-pion cloud, (b) resonance-hole 
(``rho-sobar'') excitations, and (c) the pion axialvector system.}
\label{fig:rho-self}
\end{center}
\end{figure}

\begin{figure}[!t]
\begin{center}
\includegraphics[width=0.9\textwidth]{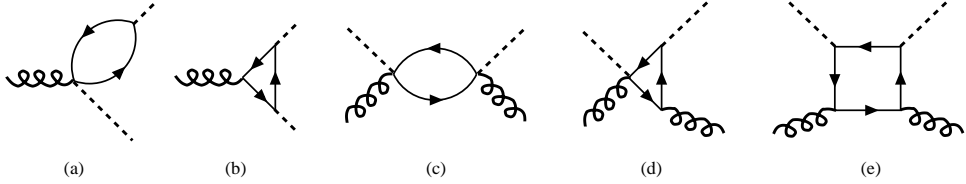}
\caption{$\rho \pi\pi$ vertex corrections, from $Nh$ and $\Delta h$ insertions
associated to the pion self-energy, accounted for in the two-pion part of the 
$\rho$ self-energy.}
\label{fig:rho-self-vertex-medium}
\end{center}
\end{figure}

For the in-medium $\rho$-meson self-energy we adopt the approach of 
Refs.~\citen{Rapp:1997ei,Urban:1998eg} which is based on a realistic 
description of the $\rho$ in free space (consistent with $\pi \pi$
$P$-wave scattering and the pion electromagnetic form-factor). In cold
nuclear matter, the self-energy is built from two components: 
$Nh$ and $\Delta h$ excitations in the two-pion 
cloud, $\Pi_{\rho\pi\pi}$, and direct excitations of baryonic resonances 
in $\rho N$ scattering, $\Pi_{\rho Bh}$, as depicted in
Figs.~\ref{fig:rho-self}\,(a) and (b). Since the 4-pion cloud, 
Fig.~\ref{fig:rho-self}\,(c), is not accounted for in the vacuum 
$\rho$ (due its much higher energy threshold), we do not include its 
nuclear modifications here\footnote{At finite temperature,
direct interactions of the $\rho$ with thermal pions can give rise to 
mesonic resonance excitations as in Fig.~\ref{fig:rho-self}\,(c), but 
they do not appear at zero temperature.}. 
These contributions have been evaluated in an effective quantum
hadro-dynamical model of meson-meson and meson-baryon interactions, 
both in cold~\cite{Urban:1998eg} and hot~\cite{Urban:1999im,Rapp:1999us} 
hadronic matter. 
For the first component, $\Pi_{\rho\pi\pi}$, special care needs to be
taken to preserve the Ward-Takahashi identities of the vector meson 
propagator, which requires a complete set of vertex 
correction diagrams, as illustrated in 
Fig.~\ref{fig:rho-self-vertex-medium}. Note that the pion-cloud
modifications have been computed at finite 3-momentum which is essential
for the applications in the present paper. 
For the second component of the $\rho$ self-energy, $\Pi_{\rho Bh}$,
a set of $\sim$10 baryonic resonances has been accounted for; 
of particular relevance is the $N^*(1520)$, which strongly couples
to $S$-wave $\rho N$ states. The parameters underlying the total 
$\rho$ self-energy have been quantitatively 
constrained by a fit of photoabsorption cross
sections on the proton and nuclei~\cite{Rapp:1997ei}, as well as
of total $\pi N\to \rho N$ cross sections.

At this point it is in order to comment on the difference in the values 
used for $\Lambda_{\pi}$ between models A and B of the pion self-energy.
The hard form-factor with $\Lambda_{\pi}\sim 1$~GeV represents a typical
value used in meson-exchange potentials of the $NN$ force where the pion
figures as a space-like degree of freedom (smaller values of 
$\Lambda_{\pi}$ can be accommodated if additional vertex effects, such as 
correlated $\pi \rho$ exchange, are incorporated~\cite{Janssen:1993nj}). 
The large cut-off value results in a strong  pion self-energy in nuclear 
matter which, in turn, needs to be tamed by short-range $NN$ correlations 
which we implement via Migdal parameters ($g'_{ij}$). To 
avoid premature pion condensation (i.e. at densities close  to saturation), the
$g'$ parameters have to be rather large ($\sim$0.8) and,  consequently, the pion
self-energy is quite sensitive to their precise  values. On the other hand, in
the $\rho$-meson self-energy, the pion self-energy figures  into the two-pion
cloud of the $\rho$. As mentioned above, this part of the in-medium $\rho$
self-energy has been 
constrained independently by relating its imaginary part in the low-density
limit to: (i) $\pi N \to \rho N$ production cross sections  (which at high
energy are  dominated by pion exchange); (ii) the non-resonant continuum in
nuclear photo-absorption (corresponding to processes of type $\gamma N \to \pi
N$, $\pi \Delta$). The first constraint puts a rather stringent limit on
$\Lambda_{\pi}$ of order $0.3$~GeV, not to overshoot the experimental cross 
section  (note that in these reactions one pion is on-shell). This situation is
reminiscent to the elastic $P$-wave $\pi N$ scattering phase shifts, where a 
$\Delta$-pole ansatz also requires a soft form-factor cut-off, $\Lambda_{\pi N
\Delta}\sim 0.3$~GeV (c.f., e.g., Refs.~\citen{Moniz:1980be,vanHees:2004vt}). 
It turns out that, with this soft form-factor, the non-resonant background in
nucleon photo-absorption [constraint (ii)] is predicted at $\sim$80~$\mu$b,
which is consistent with experiment (c.f., e.g., the discussion in  
Section~4.1.2 of Ref.~\citen{Rapp:1999ej}). The ``unnatural"  softness of the 
$\pi N N (\Delta)$ form-factor could be related to missing  cancellations 
with higher resonances, or loop corrections to the vertex in a more 
elaborate treatment of the  effective hadronic theory. Since the soft 
form-factor suppresses the in-medium ($P$-wave) pion self-energy appreciably, 
small Migdal parameters  ($\sim$0.2 or even 0) are preferred in the 
application to photo-absorption on nuclei. Recently, this model for the 
in-medium $\rho$ has been checked against data from the CLAS collaboration 
for nuclear photo-production of $\rho$ mesons and dileptons~\cite{Riek:2008ct}.
The agreement with experiment is good which corroborates the reliability of
the $\rho$-$\pi$ modeling in nuclear matter. In the present calculation of the 
in-medium $a_1$ resonance, the pion in the $\pi \rho$ cloud is probed  in very
similar kinematics as in the $\pi \pi$ cloud of the $\rho$. We therefore think
that model B is, in fact, a more realistic and consistent parameter
choice, whereas parameter set A should be considered as an `upper
limit' on the medium effects of the pion.

Finally, medium effects on the baryonic degrees of freedom of the model have
been accounted for in a schematic way. The $\rho$ self-energy
model discussed above includes such modifications in terms of parametric
in-medium widths for the nucleon and baryon resonances~\cite{Rapp:1997ei}. 
The fit of these parameters to the photo-absorption spectra results in a
small increase of the $\Delta$ decay width and larger modifications for 
higher excitations (e.g., the $N^*(1520)$ width is fitted with an in-medium 
broadening of 250~MeV at nuclear saturation density). The data provide no
compelling evidence for baryon-mass shifts in nuclei, and therefore no 
in-medium mass corrections have been introduced for the baryons (rather than 
introducing, e.g., compensating mass shifts for the nucleon and the 
$\Delta$ in $\Delta h$ excitations).
Of course, a more complete calculation would compute the baryon self-energies
microscopically, but this is beyond the scope of the present paper.
For consistency we have also incorporated these minimal (but empirically
motivated) modifications of the nucleon and $\Delta$ 
degrees of freedom into the pion self-energy in
Eqs.~(\ref{eq:piself-total},\ref{eq:piself-total-rapp}).
The effect on the pion propagator turns out to be small and
can be safely neglected from a practical point of view in the
evaluation of the $\rho\pi$ cloud self-energy of the $a_1$ (note that it
represents changes of at least second order in density).

The in-medium pion and $\rho$-meson self-energies are implemented into
the $G_{\rho\pi}$ loop function by replacing the $\rho$ and $\pi$
propagators in Eq.~(\ref{eq:G}) by their in-medium versions. The analytic
structure of $G_{\rho\pi}$ simplifies considerably by using the Lehmann
representation for the meson propagators,
\begin{eqnarray}
D_{\pi (\rho)} (q^0,\vec{q}\,;\varrho) &=& \int_{-\infty}^{\infty}
\textrm{d}\omega \, 
\frac{S_{\pi (\rho)}(\omega,\vec{q}\,;\varrho)}{q^0 - \omega +
{\rm i}\varepsilon} \ ,
\end{eqnarray}
where $S_{\pi(\rho)}(\omega,\vec{q}\,;\varrho)$ is the pion ($\rho$-meson)
spectral function,
\begin{eqnarray}
S_{\pi(\rho)}(\omega,\vec{q}\,;\varrho)=-\frac{1}{\pi} {\rm Im} D_{\pi (\rho)}
(\omega,\vec{q}\,;\varrho) \ ,
\end{eqnarray}
and
\begin{eqnarray}
D_{\pi(\rho)} (q^0,\vec{q}\,;\varrho) &=& \frac{1}{(q^0)^2 - \vec{q}\,^2 -
m_{\pi(\rho)}^2 - \Pi_{\pi(\rho)}(q^0,\vec{q}\,;\varrho)} \ .
\end{eqnarray}
After some manipulations, the $\rho\pi$ loop function in nuclear matter 
takes the form 
\begin{eqnarray}
\label{G_ITF}
G_{\rho\pi}(P^0,\vec{P}=\vec{0};\varrho) =
\int_0^{\infty} \frac{dW}{2\pi} \,
\left[ \frac{1}{P^0-W + {\rm i}\varepsilon}
 - \frac{1}{P^0+W - {\rm i}\varepsilon} \right]
 \, F(W) 	\ 
\end{eqnarray}
with
\begin{eqnarray}
\label{Fomega}
 F(W) &=& \int\limits^{q_{\rm max}} \frac{d^3q}{(2\pi)^3}
 \int_{-W}^{W} du \, \pi \
 S_{\pi}(E_+,\vec{q};\varrho) \, S_{\rho}(E_-,\vec{q};\varrho)  \ ,
\end{eqnarray}
where $E_\pm = (W \pm u)/2$. Note that 
${\rm Im}\, G_{\rho\pi}(P^0) = - F(P^0)/2$ and thus $F$ plays the role 
of a generalized in-medium two-particle phase space. 
The cut-off regularization is applied in the three-momentum integral,
Eq.~(\ref{Fomega}), and thus $F(W)$ depends explicitly on the cut-off 
parameter.

Eqs.~(\ref{G_ITF}) and (\ref{Fomega}) reduce to the vacuum expressions 
obtained in Ref.~\citen{Roca:2005nm} for vanishing nuclear density, 
$\varrho=0$. One just replaces the in-medium pion spectral function 
by $S_{\pi}(\omega,\vec{q};\varrho=0)=\delta
[\omega^2-E_{\pi}(\vec{q}\,)^2]$ and the $\rho$-meson propagator by 
$D_{\rho}(\omega,\vec{q};\varrho=0)=[\omega^2-E_{\rho}(\vec{q}\,)^2+{\rm i}\,
M_{\rho}\Gamma_{\pi\pi}]^{-1}$, which accounts for its $P$-wave 
$\pi\pi$ decay width, $\Gamma_{\pi\pi}\simeq 150$~MeV in the 
center-of-mass system (CMS). The latter expression for the $\rho$ propagator 
in vacuum neglects the real part of the $\rho$ self-energy, which is 
reabsorbed in the physical $\rho$ mass. In this work, however, consistent 
with the $\rho$ self-energy model from Refs.~\citen{Rapp:1997ei,Urban:1998eg}, 
we use the full $\rho$ spectral function,
which also accounts for the energy-dependence of the real part of the $\rho$
self-energy both in vacuum and in the medium.
For completeness, we quote the case of two stable particles in vacuum:
$F(P^0=\sqrt{s}) = (4\pi)^{-1}\, q(s)/\sqrt{s}
\,\,\theta(s-(M+m)^2)$, 
with $q(s)$ the on-shell momentum of each particle in the CMS.

It is instructive to compare our approach with earlier
calculations of the in-medium $a_1(1260)$ spectral function within 
many-body approaches. In Ref.~\citen{Kim:1999pb} the role of the excitation 
of ``meson''-sobars by the $\rho$ and $a_1$ mesons was studied. 
It was found that these mechanisms contribute to the mixing 
of the vector and axialvector correlators by inducing
extra strength in the low-energy part of the $a_1$ spectral function
(see also Ref.~\citen{Rapp:2002pn}). In the
present calculation, we do not account for ``$a_1$-sobars'' excitations
($N^*h$ excitations with the quantum numbers of the $a_1$). Nevertheless, 
as will be seen in the next section, the $a_1$ spectral function 
acquires low-energy strength from the coupling to in-medium $\rho$ and 
$\pi$ mesons. This coupling also lowers the inelastic threshold in the 
$\rho\pi$ channel all the way to $s=0$, in analogy to the 
in-medium  $\pi$ and $\rho$-meson spectral functions. 
The excitation of these low-energy modes induces a 
mixing of vector and axial correlators also in the absence of 
resonance-hole modes for the $a_1$~\cite{Krippa:1997ss,Chanfray:1999me}.
In particular, the modifications of the meson cloud can lead to a rather
pronounced non-linear density dependence of their in-medium decay widths,
especially if the vacuum threshold of the decay products is close to 
the parent's meson mass, e.g., for the  $\phi$ and 
$\omega$~\cite{Cabrera:2002hc,Cabrera:2009}. Such effects have 
recently received considerable attention in experiment and 
theory~\cite{Cabrera:2003wb,Ishikawa:2004id,Kaskulov:2006zc,Kotulla:2008xy}.

\section{$\pi\rho$ Correlations in Nuclear Matter}
\label{sec:results}

\subsection{Scattering Amplitudes}
\label{ssec:amplitude}
\begin{figure}[!t]
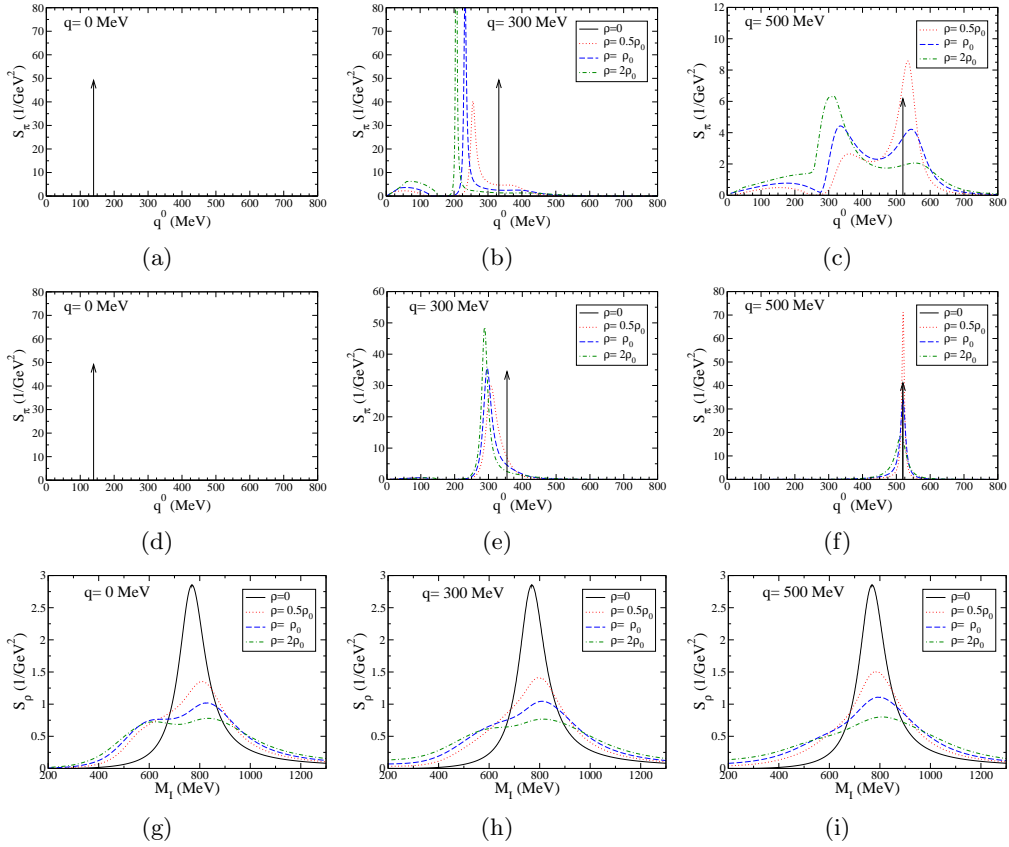

     \centering
      \subfigure[]{
          \label{fig:spectral_pi_q0}
          \includegraphics[width=.3\linewidth]{spectral_pi_q0.eps}}
      \subfigure[]{
          \label{fig:spectral_pi_q300}
          \includegraphics[width=.3\linewidth]{spectral_pi_q300.eps}}
     \subfigure[]{
          \label{fig:spectral_pi_q500}
          \includegraphics[width=.3\linewidth]{spectral_pi_q500.eps}}
      \subfigure[]{
          \label{fig:spectral_piR_q0}
          \includegraphics[width=.3\linewidth]{spectral_pi_q0.eps}}
      \subfigure[]{
          \label{fig:spectral_piR_q300}
          \includegraphics[width=.3\linewidth]{spectral_piR_q300.eps}}
     \subfigure[]{
          \label{fig:spectral_piR_q500}
          \includegraphics[width=.3\linewidth]{spectral_piR_q500.eps}}
     \subfigure[]{
          \label{fig:spectral_rho_q0}
          \includegraphics[width=.3\linewidth]{spectral_rho_q0.eps}}
     \subfigure[]{
          \label{fig:spectral_rho_q300}
          \includegraphics[width=.3\linewidth]{spectral_rho_q300.eps}}
      \subfigure[]{
          \label{fig:spectral_rho_q500}
          \includegraphics[width=.3\linewidth]{spectral_rho_q500.eps}}
  \caption{Spectral functions for the $\pi$ and $\rho$~\cite{Rapp:1999us}
   for different nuclear densities and 3-momenta as indicated in the legends.
   The first (second) row corresponds to model A (B)  for the $\pi$-self-energy.
   The $\rho$ spectral function (third row) is
   plotted as a function of $M_I\equiv\sqrt{{q^0}^2-q^2}$.}
     \label{fig:spectral}
\end{figure}

Let us start by exhibiting the previously calculated spectral functions 
for $\pi$- and $\rho$-mesons in cold nuclear matter which serve as an
input for our further calculations. Figs.~\ref{fig:spectral}(a-f) summarize
these quantities for the pion as a function of its energy 
for various 3-momenta and nuclear densities (in units of   
saturation density,  $\varrho_0=0.16\textrm{ fm}^{-3}$). As a reference,
the arrows in the upper and middle panels represent the pole position 
($\delta$-function like spectral functions) of the pion propagator in vacuum.

The pion spectral function clearly exhibits the different modes excited 
in the nuclear medium. At low momentum, the pion quasi-particle
peak carries most of the strength together with a moderate contribution of the
$Nh$ excitations at lower energies. The $\Delta h$ mode induce a sizable
attraction on the pion mode relative to free space. In model A, at momenta 
of several hundred MeV/$c$, the excitation of the $\Delta h$ becomes 
prominent and significantly broadens to the pion quasi-particle peak. 
The two models for the pion self-energy exhibit noticeable differences, 
mostly due to the different values for the form-factor cut-off parameter, 
$\Lambda_\pi$, see Eq.~(\ref{eq:fformfact}). For $\Lambda_\pi=1\gev$
(model A),  sizable strength appears at low energies (in fact, in the
space-like region, $q_0<q$) up to high momenta, arising from both
the $Nh$ mode and the quasi-pion mode. 
As density is increased (for a fixed momentum), the pion mode progressively 
approaches the $\Delta h$ energy threshold and thus exhibits a narrower 
quasi-particle structure. Using a softer form-factor, $\Lambda_\pi=0.3\gev$ 
(model B), cuts down the pion self-energy already for a few hundred MeV 
of momentum. At $q=300$\,MeV, the three-level structure of the
self-energy is less visible, and the pion is characterized by a moderately
broadened quasiparticle with considerable attraction, whereas little 
low-energy strength is induced by $Nh$ excitations. At $q=500$\,MeV the 
pion becomes a relatively narrow quasiparticle with strength concentrated 
around the free energy and a width increasing with density.

In the lower panels of Fig.~\ref{fig:spectral} 
we show the $\rho$-meson spectral function~\cite{Rapp:1999us}
as a function of $M_I\equiv\sqrt{{q^0}^2-q^2}$.
The spectral function exhibits 
a strong broadening as density increases (with a slight upward mass
shift for small 3-momentum). The shoulder structure at energies of 
around $q_0\simeq600$\,MeV is mostly due to the $N^*(1520)h$ excitations 
and collective $\Delta h$ states in the pion cloud.
This structure progressively fades away at higher momenta, 
whereas the broadening from in-medium pion-related channels
persists at moderate momenta (it also reduces at momenta 
$q\ge$1~GeV, see, e.g., Ref.~\citen{Riek:2008ct}).

\begin{figure}[tb]
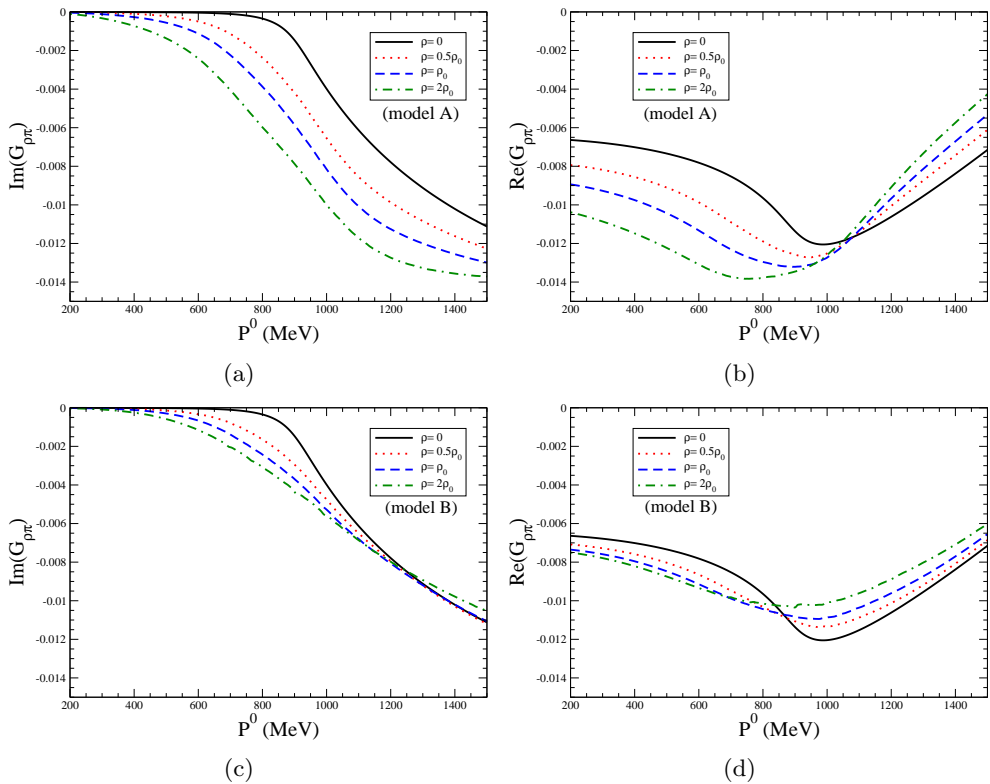

     \centering
      \subfigure[]{
          \label{fig:loop_im}
          \includegraphics[width=.45\linewidth]{res_loop_im.eps}}
      \subfigure[]{
          \label{fig:loop_re}
          \includegraphics[width=.45\linewidth]{res_loop_re.eps}}
      \subfigure[]{
          \label{fig:loop_imR}
          \includegraphics[width=.45\linewidth]{res_loop_imR.eps}}
      \subfigure[]{
          \label{fig:loop_reR}
          \includegraphics[width=.45\linewidth]{res_loop_reR.eps}}
 \caption{Imaginary (left) and real (right) parts of the $\pi \rho$ loop
 function, as a function of total energy at zero 3-momentum
 for different nuclear densities.
  The upper (lower) panels correspond to using model A (B) for the pion
    self-energy.}
     \label{fig:loop}
\end{figure}
We are now in position to inspect the consequences for the $\rho\pi$
correlations, starting with the intermediate 2-particle propagator, 
$G_{\rho\pi}(E)$, shown in Fig.~\ref{fig:loop} for different nuclear 
densities and vanishing total 3-momentum. 
The upper (lower) panels have been obtained within model A (B) of the pion
self-energy. The imaginary part,
Im\,$G_{\rho\pi}(E)$ (which we recall can 
be interpreted as a generalized $\rho\pi$ phase space at finite density),
exhibits a gradual movement of strength to lower energies with
increasing density, cf.~Fig.~\ref{fig:loop_im}; 
this strength, in particular, extends to below 
$E=3\,m_{\pi}$, which is the (absolute) vacuum threshold in this 
channel upon accounting for the $\rho$ decay width into two pions.
In Fig.~\ref{fig:loop_im} a
large increase in phase space is visible, even at the vacuum ``pole" mass 
of the $a_1$ at $E\simeq 1.1$~GeV,
which is a direct consequence of the 
excitation of low energy modes by both the $\rho$ and $\pi$
mesons in the medium, corresponding, e.g., to 
$a_1 N \to [\rho N] \to \pi\pi N$, $a_1 N \to \pi N^*(1520)$, etc.
The additional open channels increase the $a_1(1260)$ decay 
probability in the medium as compared to the vacuum.  Using 
$\pi$ self-energy B, Fig.~\ref{fig:loop_imR}, the qualitative behavior of
$G_{\rho\pi}(E)$ is similar, however the changes with density are less
pronounced, particularly around $1.2$~GeV. The reason is the use of a softer
pion form-factor, which reduces the effective phase space by cutting down the
contribution from pion-related in-medium channels at high
momenta, which are relevant at such large (total) energy.
The results for Im\,$G_{\rho\pi} = -F(E)/2$ can now be conveniently 
applied to compute the real part of $G_{\rho\pi}$ via the principle
value integral in Eq.~(\ref{G_ITF}). The results are depicted in the right
panels of Fig.~\ref{fig:loop}. One notices an increase in attraction
at energies below $800$~MeV with increasing the density, with
an accompanying increase in repulsion above $E\simeq~1$\,GeV. 
We can expect that the latter will lead to a slight increase
of the ``nominal" $a_1$ mass (or the peak of its spectral function) 
on the real energy-axis.

\begin{figure}[!t]
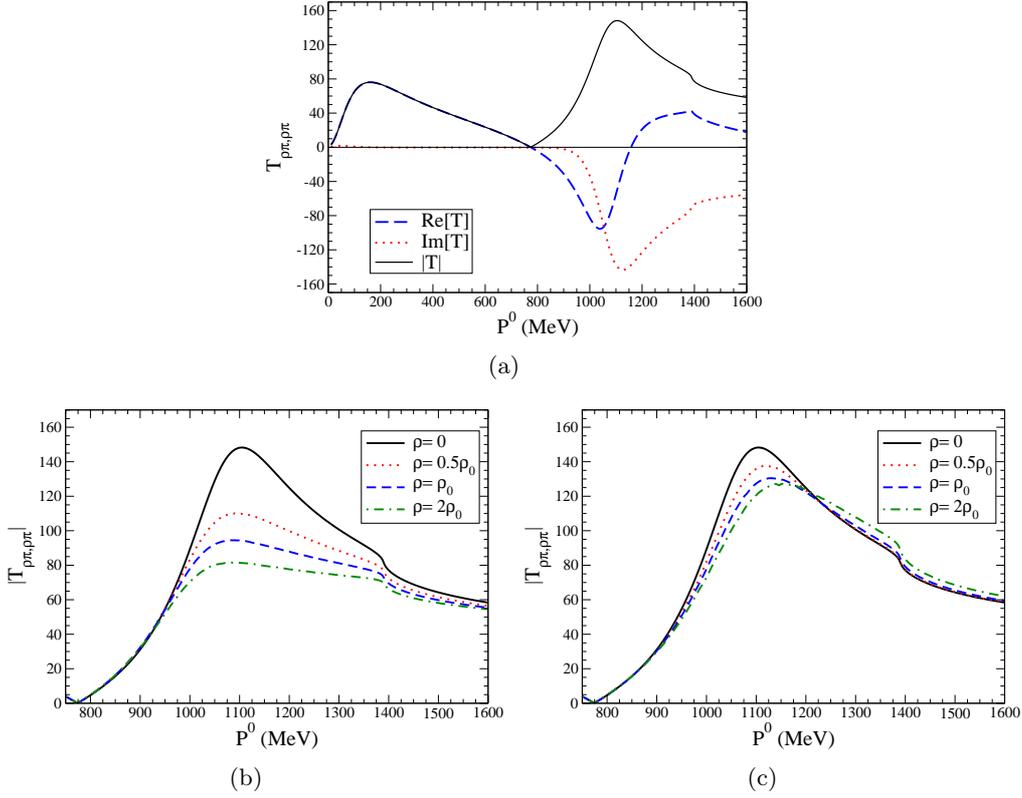

     \centering
\subfigure[]{
          \label{fig:T_re_im}
\includegraphics[width=.465\linewidth]{T_re_im.eps}}
\\
      \subfigure[]{
         \label{fig:T_rhopi}
\includegraphics[width=.465\linewidth]{T_rhopi.eps}}
     \subfigure[]{
         \label{fig:T_rhopiR}
\includegraphics[width=.465\linewidth]{T_rhopiR.eps}}
 \caption{Isospin $I=1$ $\rho\pi\to\rho\pi$ scattering amplitude.
 Upper panel: real part, imaginary part and modulus at zero density.
 Lower panels: modulus for different nuclear densities: 
 $\pi$-self-energy A, panel (b);
$\pi$-self-energy B, panel (c).}
    \label{fig:Trhopi}
\end{figure}

\begin{figure}[!t]
\begin{center}
\includegraphics[width=0.5\textwidth]{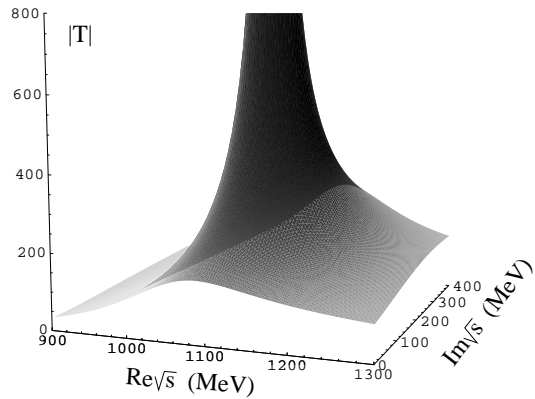}
\caption{Modulus of the $\rho\pi\to\rho\pi$ scattering amplitude in
isospin  $I=1$ in the
second Riemann sheet, showing the $a_1(1260)$ pole and its projection onto the
the real energy axis.}
\label{fig:pole3d}
\end{center}
\end{figure}

Before discussing modifications of the $\pi\rho$ scattering amplitude 
in the medium, let us first briefly comment on the result in vacuum,
cf.~Fig.~\ref{fig:T_re_im}. In this case the ``projection" of the 
$a_1(1260)$ pole onto the real axis clearly manifests itself as a 
prominent resonance structure (also note the effect of the $K\bar K^*$ 
threshold at around $E=1400\mev$). 
One of the main reasons that renders the shape of the $a_1(1260)$ 
resonance rather different from a Breit-Wigner function can be
understood  by comparing, in Fig.~\ref{fig:T_re_im}, the modulus  (solid
line), real (dashed line) and imaginary (dotted line) part of  the
$\rho\pi\to \rho\pi$ scattering amplitude.  The amplitude
 has a zero at
$775\mev$ induced by a zero in the $\rho\pi$ tree level  potential $V$,
Eq.~(\ref{eq:Vtree}), around this energy, since the full amplitude is
essentially (up to the $K^*K$ channel) proportional to $V$, recall
Eq.~(\ref{eq:bethe}). However the pole contribution itself has a
large strength at this  energy. 
This means that the non-resonant contribution (background) is as large as the pole
contribution but of opposite sign, actually canceling each other.
 As a consequence of the background,
the final shape of the amplitude is distorted
as compared to a pure pole contribution.
 All these effects are
automatically generated within our non-perturbative unitary amplitudes.
This is reminiscent to the
case of the scalar-isoscalar $\sigma(500)$ meson, where the presence of
a subthreshold Adler zero (dictated by chiral symmetry) renders its
shape rather different from a
Breit-Wigner~\cite{Oller:2004xm,Roca:2004uc}. Thus, the large distance 
of the pole from the real axis and the  presence of a zero at around
$775\mev$ are the main reasons for  deviations from a conventional
Breit-Wigner shape in  Fig.~\ref{fig:Trhopi}. For further illustration
we additionally display  in Fig.~\ref{fig:pole3d} the modulus of the
vacuum $\pi\rho\to\pi\rho$  amplitude in the second Riemann sheet as a
function of ${\rm Re}(P^0)$ and ${\rm Im}(P^0)$, which exhibits the
evolution of the  shape of $|T_{\rho\pi}|$ from the pole position to its
(distorted)  ``projection" onto the real energy axis. 

Let us now return to the nuclear medium effects of the $\rho\pi\to\rho\pi$ 
scattering amplitude, the modulus
of which is displayed for different densities in the lower panels of
Fig.~\ref{fig:Trhopi}. Fig.~\ref{fig:T_rhopi} is evaluated with
$\pi$-self-energy A and Fig.~\ref{fig:T_rhopiR} with $\pi$-self-energy B.
As density increases, the shape of the distribution around the 
$a_1(1260)$ peak widens, especially in Fig.~\ref{fig:T_rhopi}, 
following the phase-space argument given above. 
The position of the peak barely changes, again following the variations 
in the real part of $G_{\rho\pi}$, which has a moderate energy dependence
around the ``nominal" pole mass. 
Overall, the resonant structure present in free space tends to dissolve in 
the nuclear medium, which is in the spirit of the ``melting resonance'' 
scenario of chiral symmetry restoration, reminiscent to hadronic many-body 
calculations of the $\rho$ meson~\cite{Rapp:1999us,Rapp:2009yu}. 
We shall elaborate further on this assertion in the next section by studying
the first Weinberg sum rule (WSR) at finite nuclear density.

We can also evaluate the dependence of the $a_1$ pole position with the
nuclear density. The results are shown in Table~\ref{tab:poles}
for the two different pion-self-energy parameterizations, A and B.
\begin{table}[!t]
\begin{center}
\begin{tabular}{|c|c|c|}\hline 
$\varrho/\varrho_0$ & $\sqrt{s_0}$ (A) & $\sqrt{s_0}$ (B) \\ \hline
0   &   $1133-207i$ & $1133-207i$    \\ \hline
1/2 &   $1080-231i$ & $1133-220i$    \\ \hline
1   &   $1049-232i$ & $1130-235i$    \\ \hline
2   &   $1019-232i$ & $1135-254i$    \\ \hline
\end{tabular}
\end{center}
\caption{Density dependence of the $a_1$-pole position, $\sqrt{s_0}$ (units are
MeV). A and B stand for the $\pi$-self-energy model used in the calculation.}
\label{tab:poles}
\end{table}
The broadening with density visible in Fig.~\ref{fig:T_rhopi} for 
the scattering amplitude is reflected in the imaginary part of the pole 
position only up to about a density of $\varrho\simeq 0.5\varrho_0$; 
at higher densities, the suppression of $|T_{\rho\pi}|$ seems to play
an important role.
Using the $\pi$-self-energy B, the broadening with density is more
manifest while the real part remains essentially constant.
However, once again, one should be aware that the $a_1$ pole lies far
from the real axis, and therefore the correspondence between the imaginary
part of the pole position and the width of the associated distribution
(spectral function) of the resonance on the real energy axis is not
direct. 
Interestingly, the density evolution of the real part of the $a_1$ pole
indicates a slight shift to lower energies, whereas the $\rho$-meson 
spectral distribution is actually pushed to slightly 
higher energies (mostly as a consequence of two-level repulsion with
resonance-hole modes and a collective $\Delta h$ mode in the pion
cloud), cf.~Fig~\ref{fig:spectral}.
It is tempting to speculate that the mutual approach of the two poles 
(vector and axialvector) is related to a precursor of chiral
restoration, i.e., in addition to a ``melting'' of the vector and 
axialvector spectral functions, our results may suggest an additional 
trend toward ``mass degeneracy". Its effect, however, is largely
superseded by the large widths, 
$\Gamma_{\rho,a_1}^{\rm med} \gg |E_\rho^{\rm pole}-E_{a_1}^{\rm pole}|$.

\subsection{Evaluation of in-Medium Weinberg Sum Rules}
\label{ssec:WSR}
An expected precursor phenomenon of chiral symmetry restoration is a 
reduction of the pion decay constant, $f$ (an order parameter of chiral
symmetry breaking), as nuclear density increases.
Since $f$ can be interpreted as the pion-pole strength figuring into
the (longitudinal) axialvector current, its vanishing can be
realized via a degeneracy of the vector and axialvector current-current 
correlation functions if chiral symmetry is restored in the medium.
The Weinberg sum rules (WSRs) relate the pion decay constant to 
energy weighted moments of the difference between vector and axialvector
spectral functions. In particular, in vacuum, the first WSR
reads~\cite{Weinberg:1967kj}
\be
\int \frac{ds}{s}\left( \hat{S}_V-\hat{S}_A\right) = f^2
\label{eq:WSRint} \ .
\ee
In our normalization, $\hat{S}_V$ in Eq.~(\ref{eq:WSRint}) is related
to the vector spectral function, $S_V$, by
$\hat{S}_V=(sF_V/M_V)^2 S_V$.

\begin{figure}[t]
     \centering
      \subfigure[]{
          \label{fig:WSR1}
          \includegraphics[width=.28\linewidth]{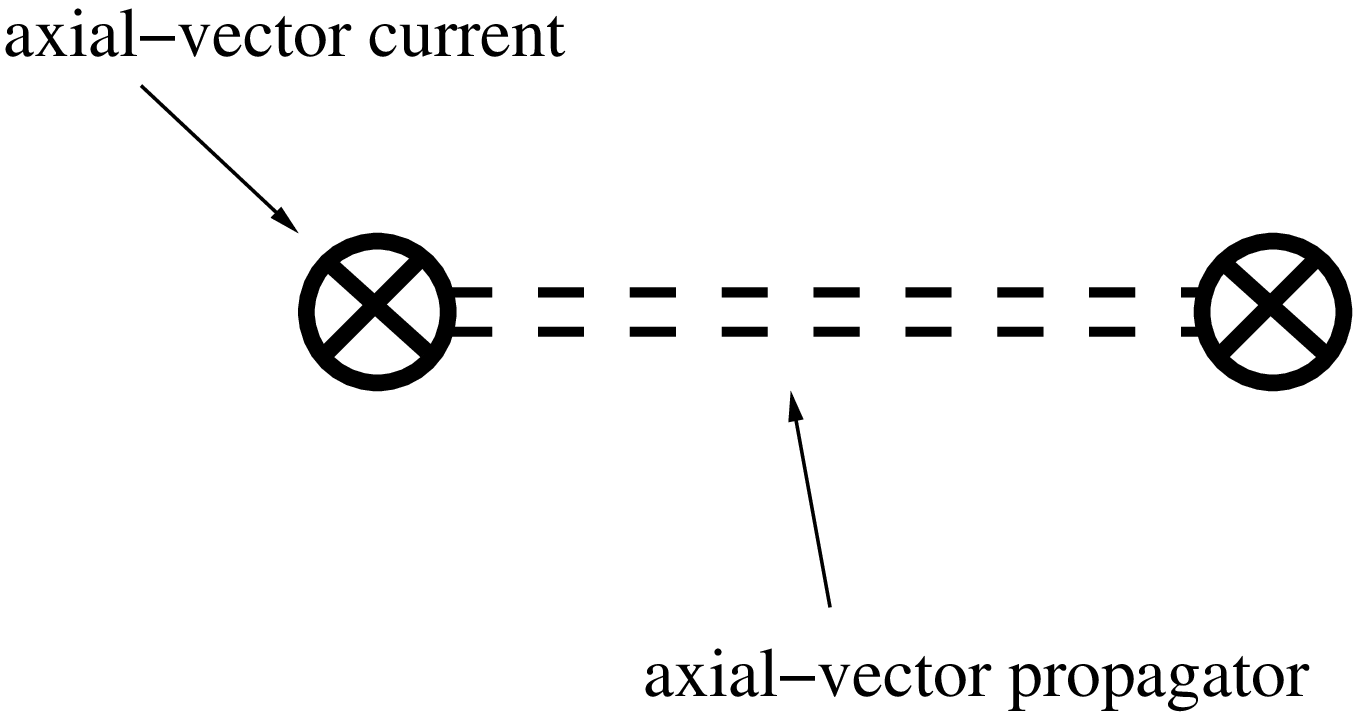}}
	  \hspace{.5cm}
      \subfigure[]{
          \label{fig:WSR2}
          \includegraphics[width=.36\linewidth]{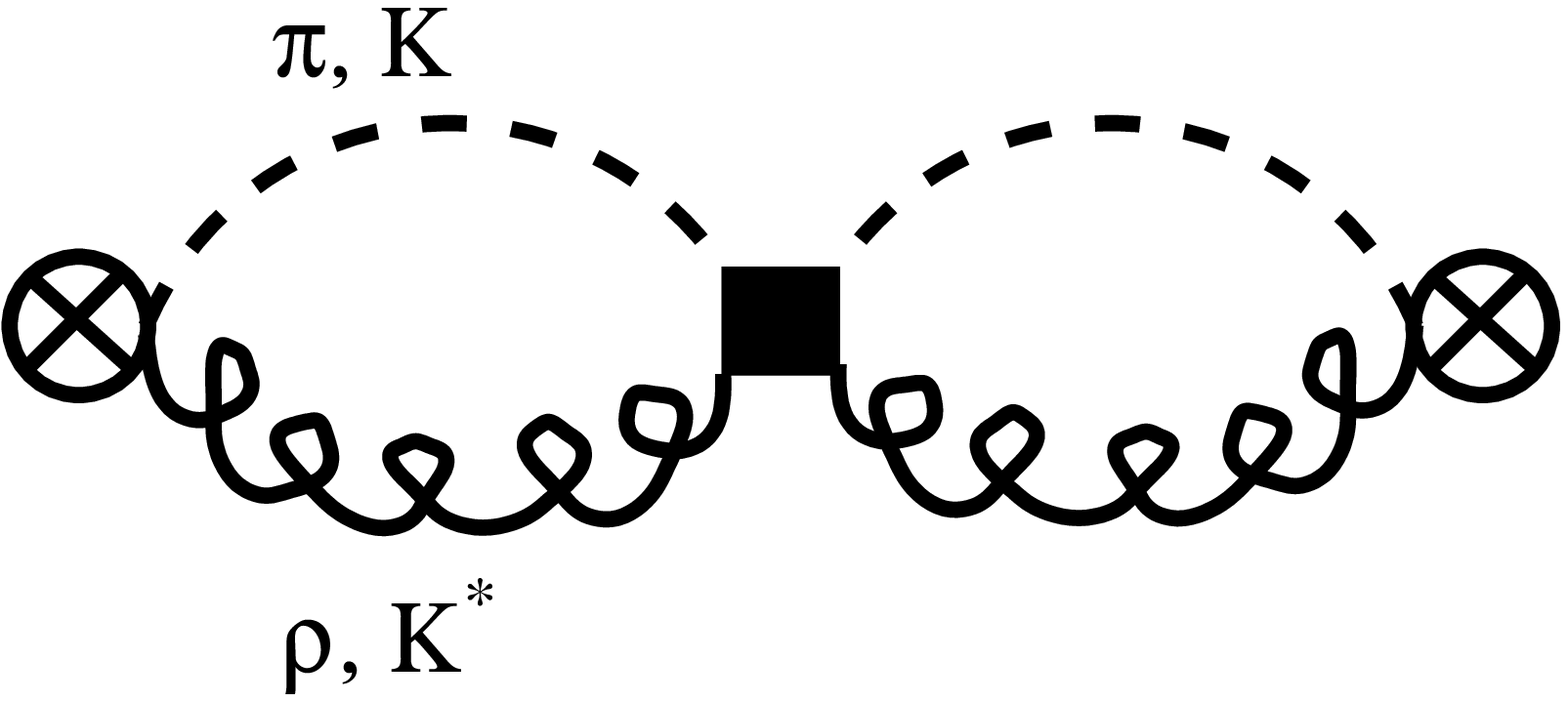}}
  \caption{Axialvector two point correlation function. 
  (a): for an explicit $a_1(1260)$ field. (b):
 for a dynamically generated  $a_1(1260)$ resonance. The square
 represents the $VP\to VP$ unitarized interaction.}
     \label{fig:WSR}
\end{figure}

The generalization of the first WSR in the medium is given in
Ref.~\cite{Kapusta:1993hq},
\be
\int_{0}^{\infty} \frac{\omega\,  d \omega}{\omega^{2}-\vec p^{\,2}}\left(
\hat{S}_V^{L}(\omega,\vec p)-\hat{S}_A^{L}(\omega, \vec p)\right) = 0 
\label{eq:WSRinMedium} \ .
\ee
Here, only the longitudinal components contribute to the sum rule,
and we work in the rest frame of the nuclear medium.
Separating out the pionic
contributions from the axialvector correlator, we consider first
the vector and axialvector meson contributions with $\vec p = 0$.
For the zero momentum limit, the longitudinal components become
degenerate with the transverse ones, which we have calculated
in the previous sections.
The low-energy vector correlator is
dominated by the contribution of the $\rho$ meson, $S_V \simeq S_{\rho}$.
Similarly, the axialvector correlator is usually considered to be
largely saturated by the axialvector meson~\cite{Donoghue:1993xb},
 as depicted diagrammatically in Fig.~\ref{fig:WSR1}.
A proper evaluation of the axialvector spectral function at all energies
requires a lengthy
calculation as well as a correct matching 
to the perturbative QCD continuum.
In Ref.~\citen{Wagner:2008gz} the low-energy part of the spectrum has been
worked out for the vacuum case, in a model calculation with dynamically 
generated axialvector mesons similar to the one presented here.
Implementing the full calculation of the axialvector spectral function
and extending it to account for nuclear medium  effects is out of the scope
of the present work.  However, we can make a simple estimation of the WSR
by using the one-pole approximation of the vector and axialvector
spectral functions.
For the pionic contributions, we work in the chiral limit ($m_\pi$=0 also
in the medium). One has to keep in mind, however, that the pion is not the 
only zero mode in the nuclear medium, but the particle-hole excitations 
with zero energy can contribute to the axialvector 
correlator~\cite{Jido:2008bk}. Thus, in general, the right-hand-side of 
the WSR in the nuclear medium is given by the summation of all 
zero-mode contributions (pion and particle-hole excitations).
It has been shown, however, that in the linear density approximation, only the
pion branch contributes to the axialvector correlator~\cite{Jido:2008bk}.
The longitudinal component of the axialvector correlator with
$\vec p = 0$ has only the $\mu\nu$=00 component, which involves the time 
component of the pion decay constant, $f_{t}$.

An approximate expression  for the WSR at zero three-momentum 
can be obtained from Eq.~(\ref{eq:WSRinMedium}) by 
considering saturation of the correlators by narrow 
 $\rho$ and $a_1$ resonances, as
\be
F_V^2-F_A^2\simeq f_{t}^2 \ ,
\label{eq:WSR}
\ee
where $F_V$ is the coupling of the vector-meson resonance to the vector current
and $F_A$ is the coupling of the axialvector resonance to the axial current.
One has to be aware that these assumptions are only fulfilled approximately
since the $a_1$ is quite broad and the axial correlator may be influenced by
non-resonant contributions. Indeed, in Ref.~\citen{Wagner:2008gz}, in a study 
of $\tau$ decays into three pions, it was shown that there is a strong 
interference of the $a_1$ resonance contribution with other non-resonant
terms in the evaluation of the vacuum axialvector spectral function.

Next we define within our approach an effective axial coupling constant,
$F_A$, estimate its density dependence and thus the density dependence 
of $f_{t}$. A similar idea was
used in Ref.~\citen{Geng:2008ag} in a study of the $N_c$-dependence
of dynamically generated axialvector resonances within the constraints of
WSR's.

The definition and normalization of the $F_A$ 
and $F_V$ couplings in our formalism are similar to the standard case when 
explicit axialvector fields, $A_\mu$, are considered in the theory:

\ba
{\cal L}_{V\phi}&=&-\frac{F_V}{\sqrt{2}M_V}<\partial_\mu V_\nu f^{\mu\nu}_+> 
\ ,
\label{eq:LFV}\\
{\cal L}_{A\phi}&=&-\frac{F_A}{\sqrt{2}M_A}<\partial_\mu A_\nu f^{\mu\nu}_->
\label{eq:LFA}
\ea
(given in an equivalent form in Ref.~\citen{Ecker:1988te} with
antisymmetric tensors for the vector meson fields) with
\ba
f^{\mu\nu}_\pm&=&u F_L^{\mu\nu}u^\dagger\pm u^\dagger F_R^{\mu\nu}u \ ,\nn \\
F_L^{\mu\nu}&=&\partial^\mu l^\nu-\partial^\nu l^\mu-i[l^\mu,l^\nu] \ ,\nn\\
F_R^{\mu\nu}&=&\partial^\mu r^\nu-\partial^\nu r^\mu-i[r^\mu,r^\nu] \ ,\nn\\
r^\mu&=&v^\mu+a^\mu \ ,\qquad
l^\mu=v^\mu-a^\mu \ .
\label{eq:defs}
\ea
We refer to Ref.~\citen{Ecker:1988te} for an explicit definition and
normalization of the different terms in Eq.~(\ref{eq:defs}).
Expanding $f^{\mu\nu}_\pm$ up to one pseudoscalar meson field, the
former Lagrangians provide the coupling of the explicit axialvector 
resonance field,
 $A_\mu$, to an axialvector current, $a_\mu$, 
through $F_A$ and the coupling of a $VP$ pair
to the same current via $F_V$. Note that the coupling of an axialvector 
current to a $VP$ pair also receives 
a contribution from the Lagrangian~\cite{Ecker:1988te}
\ba
{\cal L}=-i\frac{\sqrt{2} G_V}{M_V}<\partial_\mu V_{\nu} u^\mu u^\nu>
\ea 
with
\ba
u^\mu&=&i u^\dagger D_\mu U u^\dagger \ ,\\
D_\mu&=&\partial_\mu U-i(v_\mu+a_\mu)U+iU(v_\mu-a_\mu)
\ea
and $U=u^2$.

If we had an explicit axialvector field, one could evaluate the
two-point axialvector correlation function from the diagram in
Fig.~\ref{fig:WSR1}, using the
$F_A$ coupling in Eq.~(\ref{eq:LFA}).
Since in our scheme the axialvector resonances are dynamically 
generated they do not appear as explicit fields  in the Lagrangian.
Still, the axial resonance is linked to the axialvector current-current
correlator via a $VP$ loop, as depicted in Fig.~\ref{fig:WSR2}.
We note that the evaluation of the full
axialvector spectral function includes other non-resonant contributions
(like, for instance, the uncorrelated $3\pi$ and $\rho\pi$
continuum). However, for the WSR in the one-pole approximation,
Eq.~(\ref{eq:WSR}), the diagram in Fig.~\ref{fig:WSR2} is the 
dominant contribution around the $a_1(1260)$ pole.

Matching the contributions to the axialvector correlator from the 
diagrams in Figs.~\ref{fig:WSR1} and \ref{fig:WSR2} evaluated at 
the $a_1$ pole, we obtain
\ba
F_A&=&\frac{\sqrt{2}}{f}\frac{M_A}{s}\bigg[
 \frac{1}{M_\rho}\left(\left(\frac{F_V}{2}-G_V\right)(s+m_\rho^2-m_\pi^2)
+2G_Vm_\rho^2\right)
 G_{\rho\pi}(s)\,g_{\rho\pi}\nn\\
&& -\frac{1}{\sqrt{2}\,m_K^*}\left(\left(\frac{F_V}{2}-G_V\right)
(s+m_{K^*}^2-m_K^2)
 +2G_VM_{K^*}^2\right)
 G_{K^*K}(s)\,g_{K^*K}\bigg]  \ , \ \  \ \  \  
\label{eq:FAvsFV}
\ea
where $G_{\rho\pi}(s)$ and $G_{K^*K}(s)$ arise from the $VP$ loops  
[see Eq.~(\ref{eq:G})] in Fig.~\ref{fig:WSR2}, 
$g_{\rho\pi}$ is the effective coupling constant of
the dynamically generated $a_1(1260)$ to $\rho\pi$ for isospin $I=1$, 
and $g_{K^*K}$ the one to the $1/\sqrt{2}(|\bar K^* K>-|K^*\bar K>)$ 
negative $G$-parity state.
These couplings, including their relative phase,
are a natural output of the chiral unitary approach
and are obtained from the residues of the
$\rho\pi\to\rho\pi$ and $\rho\pi\to K^*K$ amplitudes at the
resonance pole  \cite{Roca:2005nm}.
For the numerical evaluation of Eq.~(\ref{eq:FAvsFV}), we take
$M_A=\textrm{Re}(\sqrt{s_{pole}})$. 
Eq.~(\ref{eq:FAvsFV}) provides, in general, complex values for $F_A$
since both $G_l$ and the effective couplings are generally complex.
In order to compare with the coupling, $F_A$, defined in
Eq.~(\ref{eq:LFA}) (which may have a different phase), we take in the 
following for $F_A$ in Eq.~(\ref{eq:FAvsFV}) its absolute value.

For the results presented in the former section we have used
a cut-off parameter of $q_{\rm max}\sim 1\gev$; similar results are
obtained by varying $q_{\rm max}$ around $1\gev$ within reasonable
values. However, to reproduce the empirical value of $f$, a 
quantitative numerical matching between $F_V$ and $F_A$ is required 
(the difference of $F_V^2$ and $F_A^2$ appears in Eq.~(\ref{eq:WSR})).
We do this by exploiting the latitude in $q_{\rm max}$ (our only free
parameter) and fix it at $q_{\rm max}\sim 0.835\gev$ to obtain 
$f=93\mev$ at zero density. This simple one-pole estimate shows that
it is possible to accommodate the Weinberg sum rule in vacuum 
within our approach. 

The couplings $F_V$ and $G_V$ will in general also pick up a nuclear-density 
dependence. The estimate of this dependence in our framework has large
uncertainties, mostly due to the pole approximation that we are
considering. We have checked that estimating of $F_V$ from the residue
of the $\rho$ meson pole results in a slight increase of $F_V$ with
density. However, this estimate is blurred due to the presence of the 
resonance-hole structure close to the $\rho$ mass.  Another possibility is 
to recast the $\rho$ spectral function in a two-level model, similar to what
was done in Ref~\citen{Kim:1999pb}, and then apply the narrow-resonance
limit. Again this produces a small variation in $F_V$. All these
pole-dominance estimations suffer from the fact that they ignore the
non-trivial structure of the meson spectral function on the real axis. 
With these uncertainties in mind, we have adopted a (conservative) 
estimation of the density dependence of $F_V$ of the form
$F_V=F_V^{(0)}(1\pm 0.2\varrho/\varrho_0)$, where $F_V^{(0)}$ is the vacuum
value of the coupling constant; that is, we allow for variations up to
$20$\% at normal nuclear matter densities.
Furthermore we have assumed the same amount of uncertainty
in  the value of $G_V$ to allow for possible medium modifications encoded in
this coupling.  For the numerical calculation we have used 
$F_V=156\mev$ and $G_V=69\mev$ at zero density.

Before discussing the numerical results, an explanation of the role played by
chiral symmetry in the $\pi$ self-energy model entering both the $\rho$ and
$a_1$ meson clouds is in order. Our pion self-energy is solely due to 
$P$-wave $\pi N$ interactions, operating in both isospin 
$\frac{1}{2}$ ($\pi N\to N$) and $\frac{3}{2}$ ($\pi N\to \Delta$)
channels. The isospin-averaged 
$S$-wave $\pi N$ scattering length is very small (a consequence of chiral 
symmetry). Therefore, $S$-wave pion self-energy contributions are small in 
symmetric nuclear  matter and have been neglected in our work. 
At the minimal level, the vanishing of the $\pi N$ interaction in the 
soft pion limit ($q\to 0$) is compatible with basic chiral requirements,
while the $\Delta$-nucleon-hole model is well established phenomenologically (we
note that the chiral structure of baryon resonances is still an open problem
\cite{Detar:1988kn,Jido:1998av,Jido:1999hd,Jido:2001nt}).
Regarding short-range correlations in the pion self-energy, chiral counting in
the $NN$ interaction is more involved (especially with 
one-pion exchange). In this work, when using parameter set A, we rely on 
the simplest possible 4-point interaction to remedy premature pion 
condensation. In our preferred set B, there is almost no sensitivity to
the short-range $NN$ correlations, which evades the problem of chiral
constraints in the $NN$ sector.
We emphasize that the main goal  of the Weinberg sum-rule analysis in this
section is to provide a consistent interpretation of our hadronic, many-body
description 
of the $a_{1}$ interactions with the nuclear medium in
terms of chiral symmetry restoration.
A similar line of thought has been followed in Ref.~\citen{Jido:2008bk},
where an in-medium sum rule for the quark condensate has been derived in terms
of the hadronic matrix elements of all the zero modes present in nuclear matter
(such as in-medium quasi-pions and $Nh$ excitations).
This implies that not only the pion
modes but also the nuclear many-body modes contribute to the 
modification of the quark condensate in the nuclear medium.

In Table~\ref{tab:fvsdens} we summarize results of $f$ as a function of
density for the two-$\pi$ self-energy models considered in the present
work. The
intervals shown in the ``present work'' columns represent uncertainties
from the density dependence of $F_V$ and $G_V$ as delineated above.
\begin{table}[t]
 \begin{center} 
 \begin{tabular}{|c|c|c|c|c|c|}\hline 
$\varrho/\varrho_0$ & 
present work 
& 
present work
  & ref.~\cite{Kim:1999pb} & ref.~\cite{Davesne:2000qg} &
ref.~\cite{wirzbaoller} \\ 
& ($\pi$-self-energy A) & ($\pi$-self-energy B) & & & \\
\hline 
0   & 93       &  93    & 93    & 93
&   93 \\ \hline
 1/2 & 100-108  & 91-101 & 65-78 & 87 &   81 \\ \hline
1   & 65-86    & 66-93  & 52-67 & 79 &   69 \\ \hline
 \end{tabular}
\end{center}
 \caption{Density dependence of the pion decay constant $f$, quoted
in units of MeV. The ranges in the columns designated ``present work"
represent the variation due to possible density dependences of $F_V$ 
and $G_V$.}
\label{tab:fvsdens}
 \end{table}
For comparison we show in the last three columns results from
Refs.~\citen{Kim:1999pb,Davesne:2000qg,wirzbaoller}. Our result for $f$
at $\varrho=\varrho_0$ is compatible with the general trend from previous
calculations, namely, a decrease with density, 
in agreement with the expectation from a partial restoration of
chiral symmetry in the nuclear medium.
This seems not to be the case at $\varrho=\varrho_0/2$. 
This may be due to our schematic form of the axialvector correlator
and that the WSR is estimated at the $a_1$ pole and hence it does not
account for the full structure of the meson spectral function.
We recall that a significant background (which our model provides but which
is not relevant at the pole position) is present, which, e.g., produces
a zero at about $800$~MeV in the scattering amplitude. Its interference
with other non-resonant contributions~\cite{Wagner:2008gz} notably affects 
the shape of the axialvector spectral function. Furthermore, since $f$ 
is obtained from a subtle subtraction between $F_V^2$ and
$F_A^2$ in Eq.~(\ref{eq:WSR}), small variations in the estimation of $F_A$ can
produce appreciable differences in $f$. Therefore, in order to draw more
reliable conclusions on the restoration of the chiral symmetry using the
Weinberg sum rule, a more detailed analysis beyond the one
pole approximation used here is called for, utilizing the full vector and 
axialvector spectral functions on the real axis (including  non-resonant
contributions). 
Further work in this direction, as well as an extension of the present 
calculation to finite temperatures (as relevant in the phenomenology of 
heavy-ion collisions), is in progress.

\section{Conclusions}
\label{sec:concl}
We have performed a theoretical estimation of the properties of
the $a_1(1260)$ axialvector resonance in cold nuclear matter. 
The starting point was a chiral unitary approach in which the 
vacuum $a_1(1260)$ is generated dynamically via resumming four-point 
interactions of a vector and a pseudoscalar meson in a coupled channel 
framework. In this way, the $a_1(1260)$, along with most of the low-lying
axialvector resonances, appears as a pole of the scattering amplitude in 
the complex energy plane, which is reflected by a resonant structure 
on the real (physical) energy axis. The model does not 
include the axialvector meson as an explicit field.
The underlying four-point interaction vertices are given by the lowest 
order chiral Lagrangian, with one free parameter required as a 
regularization scale for the one-loop two-particle propagator.

Medium effects on the $\rho\pi$ amplitude have been investigated  
by implementing $\rho$ and $\pi$ self-energies in cold nuclear matter, 
as obtained from well established hadronic many-body calculations.
The meson self-energies figure into the scattering amplitude through the
$VP$ loop function, thereby modifying the available phase space for the $a_1$
resonance decay due to opening of additional channels in the nuclear 
nuclear medium, e.g.,  $a_1 N \to [\rho N] \to \pi\pi N$ or 
$a_1 N \to \pi N^*(1520)$.

In the $a_1(1260)$ resonance region the $VP$ amplitude exhibits a 
broadening with increasing density, which could be indicative 
for a ``resonance melting" scenario. Even though less pronounced, we 
also found indications for an increase (decrease) of the $\rho$ ($a_1$) 
meson quasiparticle mass with density, which could suggest a tendency 
toward  ``mass degeneracy''. The relations of both mechanisms to chiral 
symmetry restoration in the vector~/~axialvector system
at finite density remain to be understood.

We have furthermore given a prescription to study the density dependence
of the pion decay constant utilizing the first Weinberg sum rule in a 
one-pole approximation. The main observations from this study are: 
(i) In the vacuum, within a chiral unitary model, the dynamically generated 
$a_1$ resonance enables to reproduce the Weinberg sum rule by adjusting
the regularization parameter of the $VP$ loop integral at ``natural'' 
magnitude; and, (ii) With increasing nuclear density, the pion decay constant 
tends to decrease, as a consequence of the ``admixture"  of in-medium
vector and pseudoscalar modes. Our analysis of the WSR is not very 
conclusive at present due to the approximations involved in our calculation.  
However, it provides a starting point for more elaborate evaluations
using the full meson spectral functions on the real energy axis.
In addition, the implementation of a chiral scheme for baryon resonance
excitations, as well as a self-consistent treatment of vector-axialvector
mixing, is required for more definite conclusions. These challenges will
be addressed in future work.

\section*{Acknowledgments}

This work is partly
supported by DGICYT contracts  FIS2006-03438, FPA2007-62777 and FPA2008-00592,
 the Fundaci\'on S\'eneca
 contract 02975/PI/05, the EU Integrated
Infrastructure Initiative Hadron Physics Project 
under Grant Agreement n.227431, 
the UCM-BSCH contract GR58/08 910309
and the Grant-in-Aid for the Global COE Program ``The Next Generation of
Physics, Spun from Universality and Emergence'' from MEXT of Japan.
D.J. acknowledges support from the 
Grant-in-Aid for Scientific Research by Monbu-Kagaku-Sho
of Japan (Nos. 20028004 and 20540273).
This work was partially done under the Yukawa International
Program for Quark-Hadron Physics.
D.C. wishes to acknowledge financial support from the ``Juan de la Cierva''
Programme (Ministerio de Ciencia e Innovaci\'on, Spain).
R.R. was supported by a U.S. National Science Foundation 
CAREER award under grant PHY-0449489.

\end{document}